\documentclass[prx,aps,floatfix,superscriptaddress]{revtex4}
\usepackage{CJK}
\usepackage{amssymb}
\usepackage{amsmath}
\usepackage{graphicx}
\usepackage[colorlinks=true,linkcolor=blue,anchorcolor=blue,citecolor=blue,urlcolor=blue]{hyperref}

\begin{document}

\title{Linear magnetoconductivity in an intrinsic topological Weyl semimetal}

\author{Song-Bo Zhang$^\dag$}
\affiliation{Department of Physics, The University of Hong Kong, Pokfulam Road, Hong Kong, China}
\thanks{These authors contributed equally to this work.}
\author{Hai-Zhou Lu$^\dag$}
\affiliation{Department of Physics, South University of Science and Technology of China, Shenzhen 518055, China}

\author{Shun-Qing Shen}\email{sshen@hku.hk}
\affiliation{Department of Physics, The University of Hong Kong, Pokfulam Road, Hong Kong, China}

\begin{abstract}
Searching for the signature of the violation of chiral charge conservation
in solids has inspired a growing passion on the magneto-transport
in topological semimetals. One of the open questions is how the conductivity
depends on magnetic fields in a semimetal phase when the Fermi energy crosses the Weyl nodes. Here, we study both
the longitudinal and transverse magnetoconductivity of a topological
Weyl semimetal near the Weyl nodes with the help of a two-node model
that includes all the topological semimetal properties. In the semimetal phase, the Fermi energy crosses only the 0th Landau bands in magnetic fields. For a finite
potential range of impurities, it is found that both the longitudinal
and transverse magnetoconductivity are positive and linear at the Weyl nodes, leading to an anisotropic and negative magnetoresistivity.
The longitudinal magnetoconductivity depends on the potential range
of impurities. The longitudinal conductivity remains finite at zero
field, even though the density of states vanishes at the Weyl nodes.
This work establishes a relation between the
linear magnetoconductivity and the intrinsic topological Weyl semimetal phase.
\end{abstract}

\maketitle


\section{Introduction}

Searching for the violation of chiral charge
conservation in solids started with Nielsen and Ninomiya's proposal
in 1983 \cite{Nielsen83plb}, in which the chiral charge is not conserved
in a 1D system of two bands with opposite chirality.
To simulate the 1D chiral bands, they proposed to use the lowest Landau
bands of a 3D semimetal, and expected that the longitudinal magnetoconductance
becomes extremely strong. Recently, thanks to the discovery of a number
of realistic materials of topological semimetals \cite{Wan11prb,Yang11prb,Burkov11prl,Xu11prl,Brahlek12prl,Wang12prb,Liu14sci,Xu15sci,Novak15prb,Weng15prx,Huang15nc,Lv15prx,Xu15sci-TaAs,Shekhar15np,Wang13prb,Liu14natmat,Neupane14nc,Jeon14natmat,Yi14srep,Borisenko14prl},
there is a growing passion on their electronic transport \cite{Liang15nmat,Feng15prbrc,He14prl,Zhao15prx,Cao15nc,Narayanan15prl,Guan15prl,Shekhar15arXiv,WangZ15arXiv}
and signatures of the chiral anomaly \cite{Hosur13physique,Kim13prl,Parameswaran14prx,Kharzeev13prb,ZhouJH15prb}.

Earlier theories on the longitudinal magnetoconductivity arrived at
various results \cite{Burkov14prl-chiral,Son13prb,Aji12prbrc,Gorbar14prb,Lu15Weyl-shortrange,Goswami15prb,Lu15Weyl-Localization,ChenCZ15arXiv}.
In the semiclassical limit, where the Landau levels are not well formed,
a positive $B^{2}$ magnetoconductivity was predicted \cite{Son13prb,Burkov14prl-chiral},
and is under intensive experimental investigation recently \cite{Li16np,ZhangCL16nc,HuangXC15prx,Xiong15sci,LiCZ15nc,LiH16nc,ZhangC15arXiv,YangXJ15arXiv}.
In the semiclassical approaches, the Fermi energy should overwhelm
the relaxation rate, not exactly at the Weyl nodes. The $B^{2}$ magnetoconductivity
is also obtained by modeling the disorder as long-range charged impurities
in the quantum limit \cite{Goswami15prb}. In the scenario similar
to that proposed by Nielsen and Ninomiya, different results have been
obtained so far, depending on models and treatments \cite{Aji12prbrc,Son13prb,Gorbar14prb,Goswami15prb,Lu15Weyl-shortrange}.
Literally, a semimetal must have a Fermi energy crossing the Weyl
nodes. Nevertheless, little attention is paid to the magnetoconduction
in the exact semimetal phase. More importantly, the Weyl nodes always
appear in pairs. The intrinsic connection of the Weyl nodes and the
inter-node scattering are two factors to affect the transport properties
because the chiral anomaly occurs between two Weyl nodes.

In this work, we start with a two-node model to investigate both
the longitudinal and transverse magnetoconductivity of a Weyl semimetal
near the Weyl nodes. The model describes a pair of Weyl nodes, and is
solvable in the presence of magnetic fields. The scattering potential
of the impurities is modeled by using a random Gaussian potential, in
which the range of potential may vary in realistic materials.
As long as the potential range is finite, we show that the longitudinal
magnetoconductivity is positive and linear in magnetic field, giving
rises to a negative magnetoresistance. As the field goes to zero,
we have a finite minimum conductivity, even though the density of
states (DOS) vanishes at the Weyl nodes. In the transverse magnetoconductivity,
we find a crossover from linear $B$ dependence in the short-range
potential limit to $1/B$ dependence in the long-range potential limit.

The paper is organized as follows. We first introduce the two-node model and present the solution of the Fermi arcs in Sec.~\ref{Sec:model}. We then discuss the magnetic field-induced DOS and chiral anomaly in Sec.~\ref{Sec:DOS}. In Secs.~\ref{Sec:Szz} and \ref{Sec:Sxx}, the formulas and analysis of the longitudinal and transverse magnetoconductivities in the presence of the random Gaussian potential are given, respectively. We also discuss the paramagnetic Weyl semimetal in Sec.~\ref{Sec:Dirac}. The conclusions are given in  Sec.\ref{Sec:conclusion}. The detailed calculations are provided in Appndix~\ref{App:DOS}-\ref{matrixelements}.
\begin{figure}[htbp]
\centering
\includegraphics[width=0.6\textwidth]{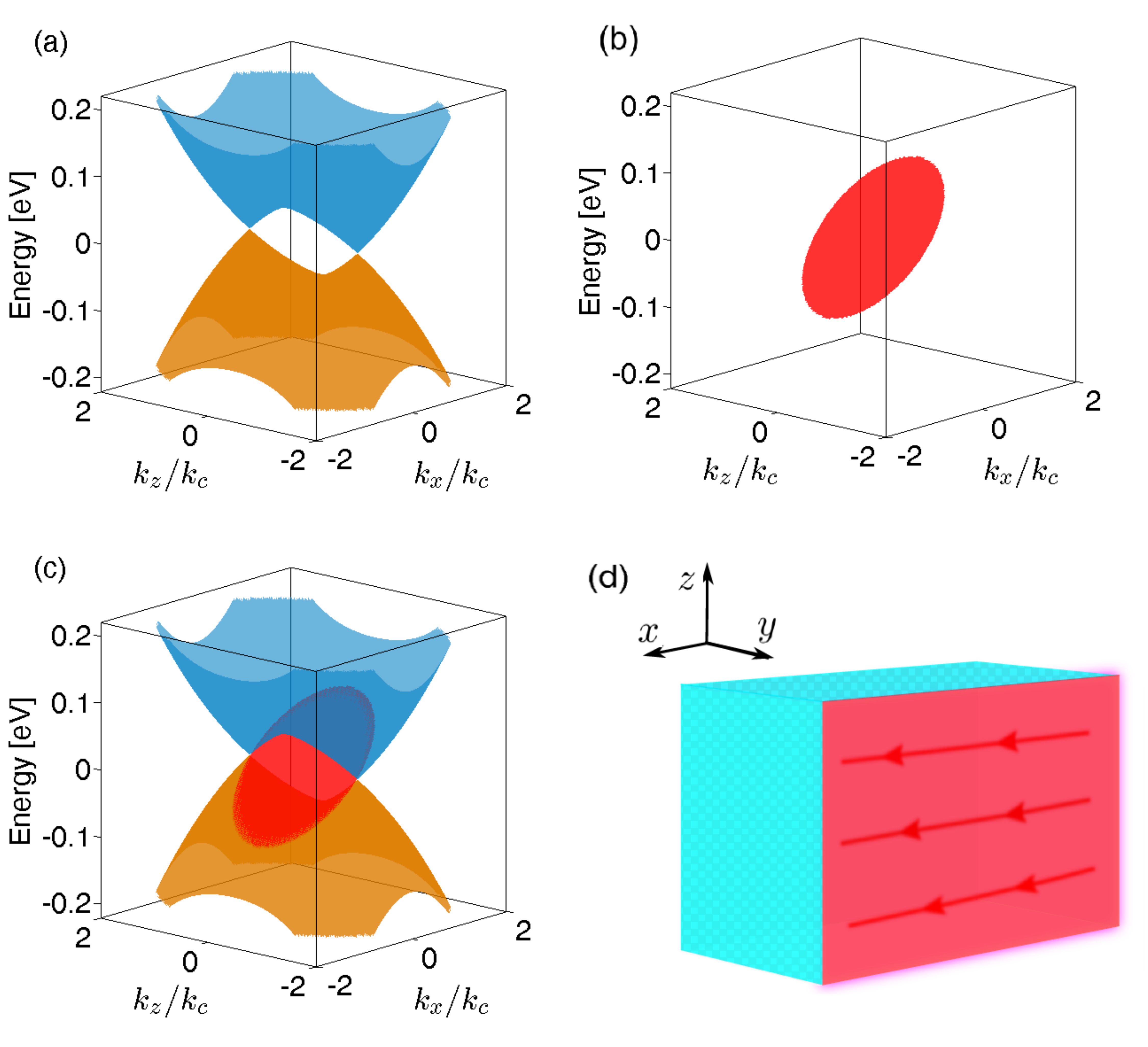}
\protect\caption{The energy spectrum of the bulk states (a) and the surface states at the $y$=$0$ surface (b) of the topological Weyl semimetal. (c) Both the bulk and surface states. (d) The real-space schematic of the topological Weyl semimetal and its surface states. We asusme that the $\hat{\bf{x}}$ and $\hat{\bf{z}}$ directions are infinitely long. The lines with arrows in the $y$=0 (red) plane indicate that the chiral surface states travel along only one direction. In contrast, a topological Dirac semimetal hosts helical surface states that travel along both the $\hat{\bf{x}}$ and $-\hat{\bf{x}}$ directions. Parameters: $k_{c}=0.1$/nm, $M=5$ eV$\cdot$nm$^{2}$, and
$A=1$ eV$\cdot$nm. }
\label{Fig:FermiArc}
\end{figure}
%

\section{Two-node model of Weyl semimetal and Fermi surfaces \label{Sec:model}}
 We describe the topological
semimetal with a two-node model \cite{Lu15Weyl-shortrange},
\begin{eqnarray}
H(\mathbf{k})=A(k_{x}\sigma_{x}+k_{y}\sigma_{y})+M(k_{c}^{2}-k_{x}^{2}-k_{y}^{2}-k_{z}^{2})\sigma_{z},\label{Eq:model}
\end{eqnarray}
where $\sigma_{x,y,z}$ are the Pauli matrices, $\mathbf{k}=(k_{x},k_{y},k_{z})$
is the wave vector, and $A$,
$M$ and $k_{c}$ are model parameters. The model is in the topological
semimetal phase and describes a pair of 3D gapless Dirac cones, with
two Weyl nodes located at $\mathbf{k}=(0,0,\pm k_{c})$ in momentum
space [see Fig. \ref{Fig:FermiArc}(a)]. The topological properties of the two-node model can be examined
by the Berry curvature, Chern number and Fermi arcs \cite{Shen12book,Lu15Weyl-shortrange}.
A topological semimetal has the $k_{z}$-dependent topologically protected
surface states for a specific $k_{z}$ between the two Weyl nodes.
This is demonstrated by a non-zero Chern number as a function of $k_{z}$,
$N_{c}(k_{z})=\mathrm{sgn}(M)$ for $\left|k_{z}\right|<k_{c}$, and
0 for $\left|k_{z}\right|>k_{c}$. According to the bulk-edge correspondence,
there exist surface (or edge) states around the surfaces parallel
to the $\hat{\mathbf{z}}$ direction. The solution of the surface states can be found from
the two-node model explicitly by following the solution
to the two-dimensional modified Dirac equation that describes the
quantized anomalous/spin Hall effects \cite{Shan11njp}.

Suppose we have a semi-infinite system in the half plane $y\leqslant0$ with open boundary conditions and with translational symmetry along the $\hat{\bf{x}}$ and $\hat{\bf{z}}$ directions, as shown by Fig.~\ref{Fig:FermiArc}(d). $k_x$ and $k_z$ are still good quantum numbers but $k_y$ is replaced $k_y= -i\partial_y$ in the Hamiltonian \eqref{Eq:model}. We can assume a trial wavefunction for each set of $k_x, k_z$ as
\begin{eqnarray}
\psi_\lambda = e^{ik_x x+ik_z z} \begin{bmatrix}
\psi_1 \\ \psi_2
\end{bmatrix} e^{\lambda y}.
\end{eqnarray}
Substituting the trial wavefunction into the eigen equation $H(k_x,-i\partial_y, k_z) \psi_\lambda =E \psi_\lambda$, we have
the secular equation for the eigen energies
\begin{eqnarray}
  \det |H(k_x,-i\lambda, k_z)-E|=0,
\end{eqnarray}
which gives four solutions of $\lambda(E)$, denoted as $\beta\lambda_\alpha$ with $\beta=\pm$, $\alpha=1,2$, and
\begin{eqnarray}
\lambda^2_\alpha(E) = & (-1)^{\alpha} \frac{\sqrt{A^4-4A^2M^2(k_c^2-k_z^2) + 4M^2 E^2}}{2M^2}
+\frac{A^2-2M^2\Delta_k }{2M^2},\label{Lambdas}
\end{eqnarray}
where $\Delta_k=k_c^2-k_x^2-k_z^2$. Each $\beta\lambda_\alpha(E)$ corresponds to a spinor state
\begin{eqnarray}
\psi_{\alpha\beta}= \begin{bmatrix}
M(\Delta_k+\lambda_\alpha^2)+E \\A(k_x+\beta\lambda_\alpha)
\end{bmatrix},\label{State1}
\end{eqnarray}
or
\begin{eqnarray}
\psi_{\alpha\beta}= \begin{bmatrix}
A(k_x-\beta\lambda_\alpha)\\
-M(\Delta_k+\lambda_\alpha^2)+E
\end{bmatrix}. \label{State2}
\end{eqnarray}
The general wavefunction in the $\hat{\bf{y}}$ direction can then be written as a superposition of the spinor states
\begin{eqnarray}
\Psi_{k_x,k_z}(E,y)= \sum_{\alpha=1,2,\beta=\pm}C_{\alpha\beta}\psi_{\alpha\beta} e^{\beta \lambda_\alpha y},
\end{eqnarray}
where the eigen-energy $E$ as well the coefficients $C_{\alpha\beta}$ are to be find from the boundary conditions.

We now apply the open boundary conditions: $\Psi_{k_x,k_z}(E,-\infty)=\Psi_{k_x,k_z}(E,0) = 0$.
The former condition $\Psi_{k_x,k_z}(E,-\infty) = 0$ requires that $\Psi$ contains only the terms with positive $\beta$ and $\mathrm{Re}(\lambda_\alpha)>0$ (or negative $\beta$ and $\mathrm{Re}(\lambda_\alpha)<0$) and thus $C_{1-}=C_{2,-}=0$. According to Eq.~\eqref{Lambdas}, this condition can only be satisfied in two cases: (i) $\lambda_{1,2}>0$, and (ii) $\lambda_{1,2}= a\mp i b$ with $a,b>0$ (Note that $\lambda_1=\lambda_2$ corresponds to a trivial case). The later condition $\Psi_{k_x,k_z}(E,0) = 0$ then gives a secular equation
\begin{eqnarray}
 \det\begin{vmatrix}
\psi_{1+} & \psi_{2+}
\end{vmatrix} =0 \label{secular}
\end{eqnarray}
to determine $C_{1+}$ and $C_{2+}$. Substituting Eqs.~\eqref{State1} and \eqref{State2} into Eq.~\eqref{secular}, respectively, and considering $\lambda_1\neq \lambda_2$, we arrive at
\begin{eqnarray}
E= & -M(\Delta_k - \lambda_1\lambda_2)+Mk_x(\lambda_1+\lambda_2), \label{SeEq1}\\
E= & M(\Delta_k - \lambda_1\lambda_2)+Mk_x(\lambda_1+\lambda_2), \label{SeEq2}
\end{eqnarray}
which lead to
\begin{eqnarray}
 \lambda_1\lambda_2 = \Delta_k . \label{Lambdasquare0}
\end{eqnarray}
On the other hand, according to Eq.~\eqref{Lambdas}, we have
\begin{eqnarray}
\lambda_1^2 \lambda_2^2 &=& ({A^2k_x^2 -E^2})/{M^2}+\Delta_k^2,  \label{Lambdasquare}\\
\lambda^2_1+\lambda_2^2 &=& {A^2}/{M^2}-2\Delta_k. \label{Lambdasquare1}
\end{eqnarray}
Substituting Eq.~\eqref{Lambdasquare0} into Eq.~\eqref{Lambdasquare}, we have immediately
\begin{eqnarray}
 E^2 =  A^2 k_x^2. \notag
\end{eqnarray}
Using Eqs.~\eqref{Lambdasquare0} and \eqref{Lambdasquare1} and keeping in mind that in both the cases (i) and (ii), $\lambda_1+\lambda_2>0$, we must have $\lambda_1+\lambda_2={A}/{|M|}>0$. Here we assume $A>0$ without loss of generality. Putting this result into Eq.~\eqref{SeEq1}, the dispersion of the surface states is finally given by
\begin{eqnarray}
E_{\text{arc}}(k_x,k_z)=\mathrm{sgn}(M)Ak_x. \label{SurfaceEnergy}
\end{eqnarray}
The corresponding wavefunction can be simplified as
\begin{eqnarray}
\Psi_{k_x,k_z}^{\text{arc}}(\mathbf{r}) & = & C e^{ik_{x}x+ik_{z}z} \begin{bmatrix}
\text{sgn}(M) \\ 1
\end{bmatrix} (e^{\lambda_1y}-e^{\lambda_2y}),
\end{eqnarray}
where $C$ is a normalization factor  and $\lambda_{1,2} =A/2|M|\mp\sqrt{(A/2M)^{2}-\Delta_k}$.
At the surface of $y=0$, the surface states are eigenstates
of $\sigma_{x}$ with a uniform effective velocity, $v_{\text{eff}}=\text{sgn}(M)A/\hbar$.
Thus they are chiral surface states around the surface parallel with
the $\hat{\mathbf{z}}$ direction. Also in both cases (i) and (ii), we have $\lambda_1\lambda_2>0$ and henceforth $\Delta_{k}>0$. Therefore the solution of Fermi surface states is restricted inside a circle defined by $k_{x}^{2}+k_{z}^{2}<k_{c}^{2}$, as shown by Figs.~\ref{Fig:FermiArc}(b) and (c).
At zero Fermi energy, i.e., $k_{x}=0$, the surface states exist for all $\left|k_{z}\right|<k_{c}$
which produces a Fermi arc connecting two Weyl nodes. For a non-zero
Fermi energy, the ends of the Fermi arc are shifted away from the
Weyl nodes until they vanish.

\section{Field-induced density of states and chiral anomaly \label{Sec:DOS}}
\begin{figure}[htbp]
\centering \includegraphics[width=0.55\textwidth]{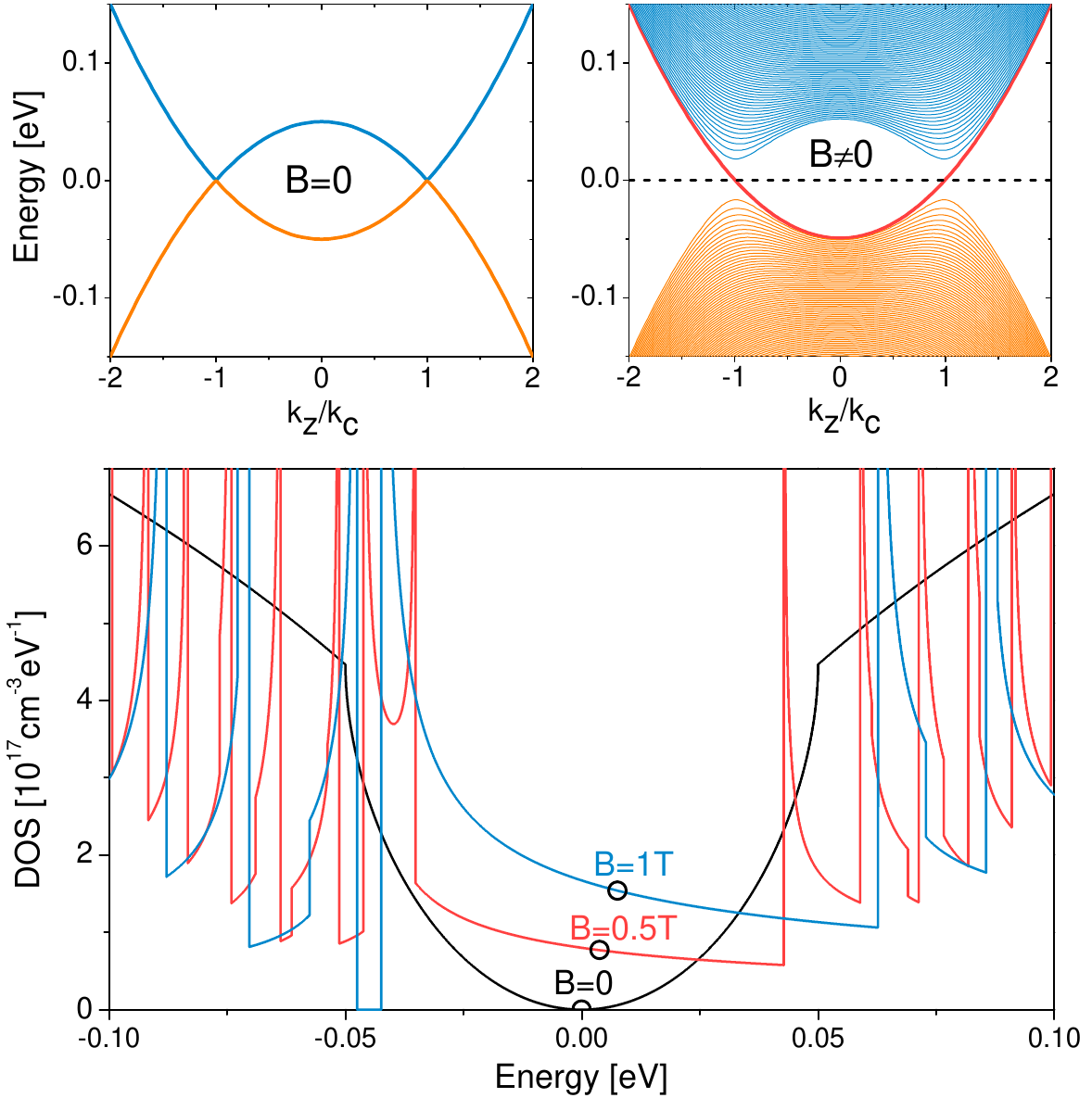} \protect\caption{Top left: the energy spectrum of a Weyl semimetal as a function of
$k_{z}$ at $k_{x}=k_{y}=0$. Top right: the Landau bands of the Weyl
semimetal in a $\hat{\bf{z}}$-direction magnetic field ($B=0.1$ T). The 0th
Landau bands are in red. Bottom: the DOS in the absence
and presence of the $\hat{\bf{z}}$-direction magnetic field. The circles indicate
the DOS at the Weyl nodes, which grows linearly with the
magnetic field. Parameters: $k_{c}=0.1$/nm, $M=5$ eV$\cdot$nm$^{2}$, and
$A=1$ eV$\cdot$nm. }
\label{Fig:DOS}
\end{figure}
After showing that the two-node model is capable of capturing all the topological
properties, now we are ready to demonstrate its transport properties
arising from its chiral properties. It is well known that the DOS of a Weyl semimetal vanishes at the Weyl nodes, following
a $E^{2}$ dependence. This can also be captured by the two-node model in which the DOS at small energy $E$ is given by
\begin{eqnarray}
N(E) = & \dfrac{E^2}{2\pi^2A^2Mk_c},
\end{eqnarray}
as shown in Fig.~\ref{Fig:DOS} (see Appendix~\ref{App:DOS} for the general formulas of the DOS). When the Fermi energy locates right at the Weyl nodes, the DOS is exactly
zero. At this exact semimetal phase, it is expected to have exotic
electronic transports at zero temperature. It is found that a finite DOS can be generated by a magnetic field $B$ along the direction that connects the two Weyl nodes, i.e., a $\hat{\mathbf{z}}$-direction magnetic field in the present work. The $\hat{\mathbf{z}}$-direction
magnetic field can split the Dirac cones into a bundle of 1D bands
of Landau levels dispersing with $k_{z}$ \cite{Jeon14natmat,Lu15Weyl-shortrange,Shen04prl},
as shown in Fig.~\ref{Fig:DOS}. Figure~\ref{Fig:DOS} shows how the DOS in the absence of the magnetic field evolves
into a set of diverging peaks in the presence of the magnetic field.
The divergence is due to the van Hove singularity at the band edges
of the 1D Landau bands.

In the presence of a magnetic field in the $\hat{\mathbf{z}}$ direction, the energies of electrons form a set of Landau bands dispersing with $k_{z}$ as following \cite{Lu15Weyl-shortrange}
\begin{eqnarray}\label{Energy-LL}
E^{\nu\pm}_{k_{z}} & = & \omega/2\pm\sqrt{\mathcal{M}_{\nu}^{2}+\nu\eta^{2}},\ \nu\ge1\nonumber \\
E^{0}_{k_{z}} & = & \omega/2-Mk_c^2+Mk_{z}^{2},\ \ \nu=0
\end{eqnarray}
where $\mathcal{M}_{\nu}=Mk_c^2-Mk_{z}^{2}-\omega\nu$, $\eta=\sqrt{2}A/\ell_B$ and $\omega=2M/\ell_B^2$. $\ell_{B}\equiv\sqrt{\hbar/|eB|}$ is the magnetic length. Each Landau band has the Landau degeneracy $N_{L}=1/2\pi\ell_{B}^{2}=|eB|/h$ in a unit area in the x-y plane.

In the magnetic field, the DOS near the Weyl nodes is given by
\begin{eqnarray}
N(E) = & \dfrac{1}{4\pi^2 \ell_B^2} \dfrac{1}{\sqrt{M}\sqrt{E+Mk_c^2-\omega/2}},
\end{eqnarray}
which is contributed by the 0th Landau bands.
At the Weyl nodes, the DOS reduces to $N_{k_{c}}=(1/2\pi Mk_{c})\times(1/2\pi\ell_{B}^{2})$,
which grows linearly with the Landau degeneracy $N_L=|eB|/h$
and hence the magnetic field $B$.
This resulting DOS leads to a semimetal to metal transition,
which is under intensive investigation in the context of the disorder-induced
DOS \cite{Fradkin86prb,Shindou10njp,Goswami11prl,Ominato14prb,Syzranov15prl},
instead of in the presence of magnetic field.

In the $\hat{\mathbf{z}}$-direction field, the Fermi energy crosses
with the 0th Landau bands at $k_{c}$ and $-k_{c}$, where the Fermi
velocities are opposite, i.e., $v_{F}=\pm2Mk_{c}/\hbar$. This chiral property
of the 0th bands is exactly the scenario required in Nielsen and Ninomiya's
proposal for the chiral anomaly \cite{Nielsen83plb}. As an electric
field is applied along the Weyl node direction, the changing rates
of charge carriers are $dN_{\pm}/dt=\pm(ev_{F}/2\pi)E$ near the two Weyl
nodes at $k_{z}=\pm k_{c}$, respectively. Thus charges can be pumped from near
$k_{c}$ to $-k_{c}$, literally leading to the non-conservation of
chiral charge, i.e., the chiral anomaly \cite{Hosur13physique}. Later,
we will focus on the electronic transport in this situation when the
chiral anomaly happens. In the following discussions, we will constrain
to the case near the Weyl nodes, i.e., the Fermi energy is located
between two Landau bands of $1\pm$, and at very low temperatures,
$k_{B}T\ll \omega,\eta$.

\section{Longitudinal magnetoconductivity \label{Sec:Szz}}
At sufficiently low temperatures,
i.e., $k_{B}T\ll \omega,\eta$, and not far away from the Weyl nodes,
the electronic transport can be effectively conducted by the 0th bands
of Landau levels. When the electric and magnetic fields are in parallel with
each other, the changing rate of density of charge carriers near one
node is maximal according to the picture of chiral anomaly. In this
case, the semiclassical conductivity of the 0th Landau bands can be
found with the help of the standard Green function formulism \cite{Lu15Weyl-shortrange}. Alternatively,
it can be simply figured out by using the Einstein relation $\sigma_{zz}=e^{2}N_{F}D,$
where the DOS can be found as the Landau degeneracy
times the DOS of one-dimensional systems, i.e., $N_{F}=(1/2\pi\ell_{B}^{2})\times(1/\pi\hbar v_{F})$.
$D=v_{F}^{2}\tau^{0,\text{tr}}_{k_F}$ is the diffusion coefficient in one dimension.  $\tau^{0,\text{tr}}_{k_F}$ is the transport time, $v_F$ is the Fermi velocity in the $\hat{\bf{z}}$ direction and $k_F=\hbar v_F/2M$ is the Fermi wave number. For the scattering among the states on the Fermi surface of the $0$th Landau bands, the transport time can be found as
\begin{eqnarray}
\frac{\hbar}{\tau^{0,\text{tr}}_{k_F}}
&=& 2\pi \sum_{k_x',k_z'}
\langle |U^{0,0}_{k_x,k_F;k_x',k_z'}|^2 \rangle \delta(E_F-E^0_{k_z'}) (1-\frac{v^z_{0,k_z'}}{v_F}),
\end{eqnarray}
where $U^{0,0}_{k_x,k_F;k_x',k_z'}$ represents the scattering matrix elements and $\langle ... \rangle$ means the impurity average (see Appendix~\ref{matrixelements} for details).

The transport time $\tau^{0,tr}_{k_F}$ is sensitive to the scattering potential in materials. One of the
convenient choices is the random Gaussian potential
\begin{eqnarray}
U(\mathbf{r}) & = & \sum_{i}\frac{u_{i}}{(d\sqrt{2\pi})^{3}}e^{-|\mathbf{r}-\mathbf{R}_{i}|^{2}/2d^{2}},\label{U-Gaussian}
\end{eqnarray}
where $u_{i}$ measures the scattering strength of a randomly distributed impurity at $\mathbf{R}_{i}$, and $d$ is a parameter that determines the range of the scattering potential.
The Gaussian potential allows us to study the effect of the potential
range in a controllable way, which we find it crucial in the present
study. Now we have two characteristic lengths, the potential range
$d$ and the magnetic length $\ell_{B}$, which define two regimes,
the long-range potential regime $d\gg\ell_{B}$ and the short-range
potential limit $d\ll\ell_{B}$. Note that, for a given $d$ in realistic
materials, varying the magnetic field alone can cross between the
two regimes. Empirically, the magnetic length $\ell_{B}$ = 25.6 nm
/$\sqrt{|B|}$ with $B$ in Tesla. In the strong-field limit, e.g.,
$B>10$ T, the magnetic length $\ell_{B}$ becomes less than 10 nm,
it is reasonable to regard smooth fluctuations in materials as long-range.
\begin{figure}
\centering \includegraphics[width=0.55\textwidth]{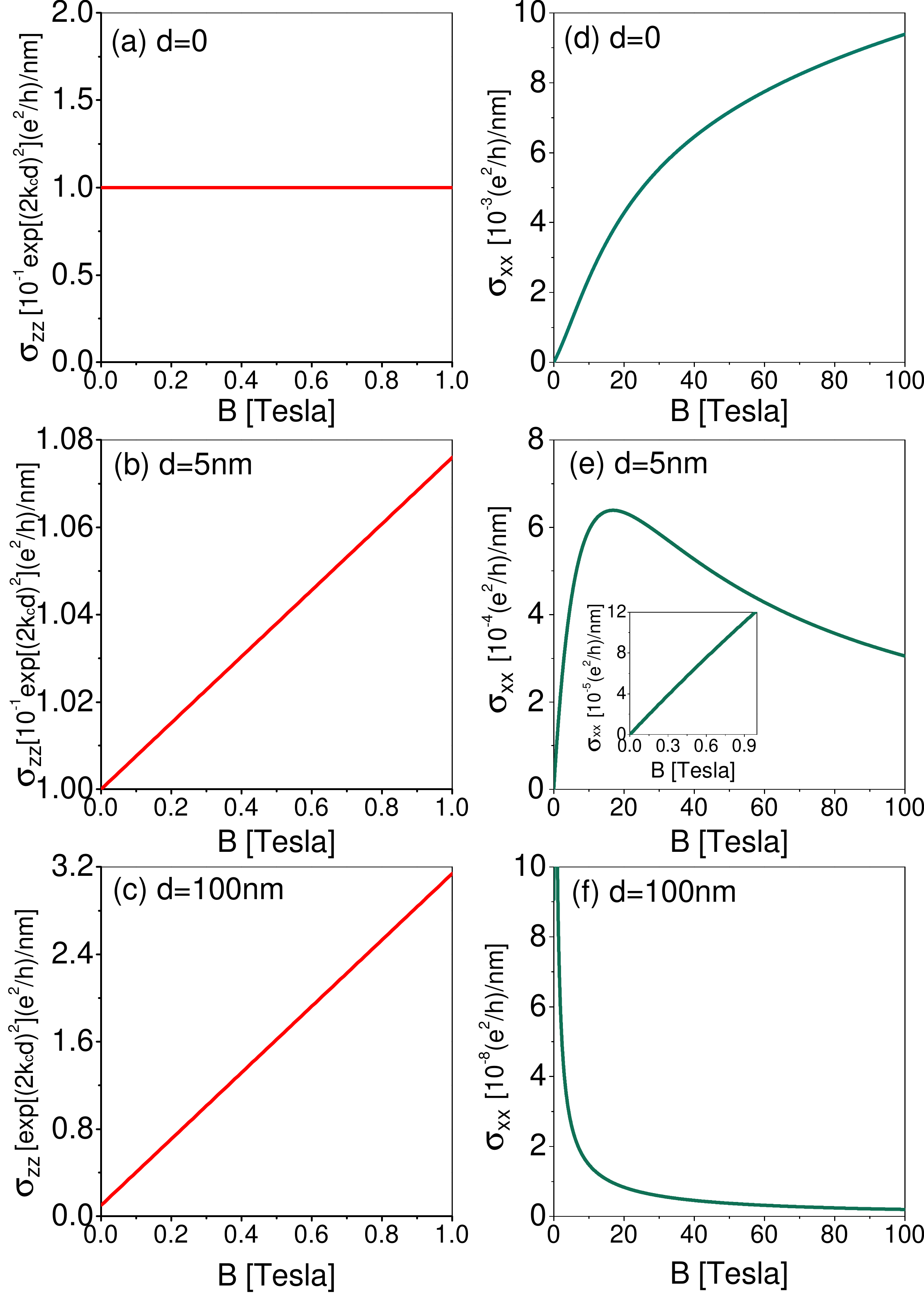}
\protect\caption{The longitudinal conductivity $\sigma_{zz}$ and transverse conductivity
$\sigma_{xx}$ of the Weyl semimetal in the $\hat{\bf{z}}$-direction magnetic
field $B$ for different potential ranges. The shared parameters: $k_{c}=0.1$/nm,
$M=5$ eV$\cdot$nm$^{2}$, $A=1$ eV$\cdot$nm, $V_{\text{imp}}=10$ (eV)$^{2}$$\cdot$nm$^{3}$. }
\label{Fig:Weyl}
\end{figure}

With the random Gaussian potential, we can find the transport time as well as the conductivity. In particular, at the Weyl nodes the transport time is obtained as (see Appendix~\ref{sec:tau-0tr} for details)
\begin{eqnarray}\label{tau-0-gaussian-twonode-B}
\frac{\hbar}{\tau^{0,\text{tr}}_{k_c}}
&=& \frac{V_{\text{imp}}}{2\pi Mk_c}
\frac{ e^{- 4d^2 k_c^2}}{2d^2+\ell_B^2},
\end{eqnarray}
and hence the longitudinal conductivity
\begin{eqnarray}
\sigma_{zz}(B) & = & \frac{e^{2}}{h}\frac{(2Mk_{c})^{2}( 2d^{2}+\ell_{B}^{2})}{V_{\text{imp}}\ell_{B}^{2} }e^{4 d^{2}k_{c}^{2}}, \label{sigma-zz-gaussian-vF}
\end{eqnarray}
where $V_{\text{imp}}\equiv \sum_i u_i^2/V $ measures the strength of the scattering and $V= L_xL_yL_z$ is the volume of the system. $L_{x,y,z}$ are the sizes of the system along the $\hat{\mathbf{x}}$, $\hat{\mathbf{y}}$ and $\hat{\mathbf{z}}$ directions, respectively.  This conductivity is generated by the inter-node scattering with a momentum
transfer of $2k_{c}$. As the magnetic field goes to zero, the magnetic length diverges and  $d/\ell_B\rightarrow 0$, and Eq. (\ref{sigma-zz-gaussian-vF}) gives a minimum conductivity
\begin{eqnarray}
\sigma_{zz}(0) & = & \frac{e^{2}}{h}\frac{ 4(Mk_{c})^{2}}{V_{\text{imp}}  }e^{4d^{2}k_{c}^{2}},\label{sigma-zz-gaussian-minimum}
\end{eqnarray}
even though the DOS vanishes  at the Weyl nodes at zero
magnetic field. A similar result was found in the absence of the Landau
levels \cite{Ominato14prb}.

According to $d$, we have two cases. (1) In the short-range limit,
$d=0$, then $\sigma_{zz}$ does not depend on the magnetic field,
giving a zero magnetoconductivity, which recovers the result for the
delta potential \cite{Lu15Weyl-shortrange,Goswami15prb}. (2) As long
as the potential range is finite, i.e., $d>0$, we can have a magnetoconductivity.
Using Eq.~(\ref{sigma-zz-gaussian-vF}),
\begin{eqnarray}
\Delta\sigma_{zz}(B) & \equiv & \dfrac{\sigma_{zz}(B)-\sigma_{zz}(0)}{\sigma_{zz}(0)}=\dfrac{B}{B_{0}},\label{sigma-zz-gaussian-B}
\end{eqnarray}
where $B_{0}=\hbar/2ed^{2}$. Thus the magnetoconductivity is given
by the range of impurity potential, and independent of the
model parameters. This means that we have a positive linear $\hat{\mathbf{z}}$-direction
magnetoconductivity for the Weyl semimetal. A finite
carrier density $n_{0}$ can drive the system away from the Weyl nodes,
then $k_{c}$ in Eq.~(\ref{sigma-zz-gaussian-vF}) is to be replaced
by $k_{F}=k_{c}+\mathrm{sgn}(M)2\pi^{2}\ell_{B}^{2}n_{0}$. The finite
$n_{0}$ can vary the linear-$B$ dependence, but a strong magnetic
field can always squeeze the Fermi energy to $k_{c}$, and recover
the linear magnetoconductivity.

A linear-$B$ magnetoconductivity arising from the Landau degeneracy has been obtained before \cite{Son13prb,Gorbar14prb}, based on the assumption that the transport time and Fermi velocity are constant. However, in the present case, we have taken into account the magnetic field dependence of the transport time, and thus the $B$-linear magnetoconductivity here has a different mechanism as a result of the interplay of the Landau degeneracy and impurity scattering. Also, in the presence of the charged impurities, a $B^{2}$ magnetoconductivity can be found in the quantum limit \cite{Goswami15prb}. A $B^2$ magnetoconductivity can also be found in the semiclassical
limit \cite{Son13prb,Burkov14prl-chiral}.

\section{Transverse magnetoconductivity \label{Sec:Sxx}}
 When electric and magnetic
fields are perpendicular to each other, the changing rate of density
of charge carriers near each node vanishes. In this case, because
the Landau bands in the $\hat{\mathbf{z}}$-direction magnetic field only disperse
with $k_{z}$, the effective velocity along the $\hat{\bf{x}}$ direction $v_{x}=\partial E_0/\hbar \partial k_{x}=0$.
The leading-order $\hat{\bf{x}}$-direction conductivity arises from the inter-band
velocity and the scatterings between the 0th bands with the bands of $1\pm$,
which are higher-order perturbation processes. Thus the transverse conductivity is usually much smaller than the longitudinal conductivity.

When the Fermi energy crosses only the $0$th bands, the leading-order conductivity is given by \cite{Datta1997}
\begin{eqnarray}\label{sigma-xx-nu-def}
\sigma_{xx}(B) &=& \sigma_{xx}^{+}(B) + \sigma_{xx}^{-}(B), \notag \\
\sigma_{xx}^{\pm}(B) &=&\frac{e^2\hbar}{\pi V}\sum_{k_x,k_z}
\text{Re}\left( G^R_{0,k_z} v^x_{0,1\pm}G^A_{1\pm,k_z} v^{x}_{1\pm,0} \right),
\end{eqnarray}
where $G_{0,k_z}^{R}=1/(E_{F}-E_{k_{z}}^{0}+i\hbar/2\tau^{0}_{k_z})$ is the retarded Green's function of electrons in the 0th bands, $G^A_{1\pm,k_z} ={1}/({E_F\mp E^{1\pm}_{k_z} - i\hbar/2\tau^{1\pm}_{k_z}})$ are the advanced Green functions of electrons in the $1\pm $ bands, respectively. $\tau^{0}_{k_z}$ and $\tau^{1\pm}_{k_z}$ are the corresponding lifetimes. The inter-band velocities along the $\hat{\bf{x}}$ direction ${v}^x_{0,1\pm}$ are given by
\begin{eqnarray} \label{v_x10}
{v}^x_{0,1+}
&=& \frac{\ell_B}{\sqrt{2}\hbar}\Big(\eta \cos \frac{\theta^{1}_{k_z}}{2}+\omega \sin \frac{\theta^{1}_{k_z}}{2}\Big), \label{v_x10-plus} \\
{v}^x_{0,1-}
&=& \frac{\ell_B}{\sqrt{2}\hbar}\Big(\eta\sin\frac{\theta^{1}_{k_z}}{2}-\omega\cos \frac{\theta^{1}_{k_z}}{2}\Big), \label{v_x10-minus},
\end{eqnarray}
where $\cos\theta^1_{k_z}=\mathcal{M}_{1}/\sqrt{\mathcal{M}_{1}^{2}+\eta^{2}}$. Considering that Fermi energy crosses only the $0$th bands and that $\hbar/\tau^0_{k_z}$ and ${\hbar}/{\tau^{1\pm}_{k_F}}$ are very small, Eq.~\eqref{sigma-xx-nu-def} can be simplified to
\begin{eqnarray}
\sigma_{xx}^{\pm}(B)
&=&\frac{e^2\hbar}{  V} \sum_{k_x,k_z} |v^x_{0,1\pm}|^2
\frac{\delta(E_F-E^0_{k_z})}{ (E_F-E^{1\pm}_{k_z})^2}\dfrac{\hbar}{2\tau_{k_z}^{1\pm}}.  \label{sigma10}
\end{eqnarray}
%


%
%
%
%

The lifetimes due to the random Gaussian potential can be calculated by using the Born approximation,
\begin{eqnarray}
\frac{\hbar}{\tau_{k_F}^{1\pm}} &=& 2\pi \sum_{k_x',k_z'}
\langle |U^{1\pm,0}_{k_x, k_F;k_x',k_z'}|^2 \rangle \delta(E_F-E^0_{k_z'}).
\end{eqnarray}
After lengthy but straightforward calculations, they can be obtained as (see Appendix~\ref{sec:tau-0} for details)
\begin{eqnarray}\label{tau-1}
\frac{\hbar}{\tau^{1\pm}_{k_F}}
&=&\frac{V_{\text{imp}}\ell_B^2}{16\pi\hbar v_F} \frac{1 +e^{- 4 d^2 k_F^2}}{(d^2+\ell_B^2/2)^2}(1\mp\cos{\theta^1_{k_F}}). \label{tau1plus}
\end{eqnarray}
The substitutions of Eqs.~\eqref{v_x10-plus} and \eqref{v_x10-minus} for $v_{0,1\pm}^x$ and of the results \eqref{tau-1} for $\tau^{1\pm}_{k_F}$ into Eq.~\eqref{sigma10} give
\begin{eqnarray}
\sigma_{xx} (B) & = & \frac{e^{2}}{h}\frac{V_{\text{imp}} }{4\pi^{2}v_{F}^{2}}\frac{\ell_{B}^{2}(1+e^{-4d^{2}k_{F}^{2}})}{(2d^{2}+\ell_{B}^{2})^{2}}\mathcal{F}(k_F),
\end{eqnarray}
where
\begin{eqnarray}
\mathcal{F}(k_z)=&\dfrac{1}{\ell_B^2} \Bigg[
\dfrac{(\hbar v_{0,1+}^x)^2\sin^2\frac{\theta^1_{k_z}}{2}}{\big(E_{k_z}^0-E_{k_z}^{1,+}\big)^2}  + \dfrac{(\hbar v_{0,1-}^x)^2 \cos^2\frac{\theta^1_{k_z}}{2}}{\big(E_{k_z}^0-E_{k_z}^{1,-}\big)^2}
\Bigg]. \label{funcF}
\end{eqnarray}

When the Fermi energy is at the Weyl node,
 the transverse conductivity $\sigma_{xx}$ at the Weyl nodes is given by
\begin{eqnarray}
\sigma_{xx}(B) & = & \frac{e^{2}}{h}\frac{V_{\text{imp}}}{16\pi^{2}M^{2}k_{c}^{2}}\frac{\ell_{B}^{2}(1+e^{-4d^{2}k_{c}^{2}})}{(2d^{2}+\ell_{B}^{2})^{2}}\mathcal{F},
\end{eqnarray}
where $\mathcal{F}=(2\omega^2+\eta^2)/(4\omega^2+4\eta^2)$, so $\mathcal{F}\approx 1/2$ in the strong-field limit ($\ell_{B}\ll M/A$)
and $\mathcal{F}\approx1/4$ in the weak-field limit ($\ell_{B}\gg M/A$).
We choose the value of $V_{\text{imp}}$ so that the band broadening is always much weaker than the spacing between the 0th and 1st Landau bands for an arbitrary magnetic field. It is safe to assume that only the 0th Landau band contributes to the conductivity at half filling.
There are three cases as shown in Figs.~\ref{Fig:Weyl} (d)-(f). At
$d=0$, $\sigma_{xx}$ reduces to the result for the delta potential
and $\sigma_{xx}\propto B$, a linear magnetoconductivity as $\sigma_{zz}$,
but much smaller \cite{Lu15Weyl-shortrange}. In the long-range potential
limit $d\gg\ell_{B}$, we have $\sigma_{xx}\sim1/B$, which gives
a negative magnetoconductivity. For a finite potential range $d$,
we would have a crossover of $\sigma_{xx}$ from $B$-linear to $1/B$
dependence. Alternatively, as shown in Fig.~\ref{Fig:Weyl} (e), for
a finite $d$ ($=5$ nm) comparable to the magnetic length $\ell_{B}$,
we have a crossover of $\sigma_{xx}$ from a linear-$B$ dependence
in weak fields to a $1/B$ dependence in strong fields. While at $d=0$
and $d\gg \ell_B$, we have the two limits as shown in Figs. \ref{Fig:Weyl} (d) and (f), respectively. For shorter $d$, a larger critical magnetic field for the crossover is needed. Figure \ref{Fig:Weyl}
also shows that the conductivity is larger for shorter $d$, so the $1/B$ transverse magnetoconductivity in the long-range limit may not survive when there are additional short-range scatters.

In particular, in Fig.~\ref{Fig:Weyl} (f), $\sigma_{xx}\propto1/B$
in the long-range potential limit. In the field perpendicular to the
$x\text{-}y$ plane, there is also a Hall conductivity $\sigma_{yx}=\mathrm{sgn}(M)(k_{c}/\pi)e^{2}/h+en_{0}/B$,
where the first term is the anomalous Hall conductivity and the second
term is the classical conductivity. In weak fields, the classical Hall
effect dominates, then both $\sigma_{xx}$ and $\sigma_{yx}$ are
proportional to $1/B$, and the resistivity $\rho_{xx}=\sigma_{xx}/(\sigma_{xx}^{2}+\sigma_{yx}^{2})$
is found to be linear in $B$. Note that here the linear MR in perpendicular fields has a different scenario compared to the previous works \cite{Abrikosov98prb,Song15prb}. Abrikosov used the Hamiltonian $v \vec{k}\cdot\vec{\sigma}$
with linear dispersion and modelled the disorder by the screened Coulomb potential under the random phase approximation \cite{Abrikosov98prb}. Song \emph{et al}. discussed a semiclassical mechanism \cite{Song15prb}.

\section{Paramagnetic Weyl semimetal \label{Sec:Dirac}}
 The existence of the non-zero
Chern number or chiral surface states indicates the time reversal
symmetry breaking in the two-node model in Eq.~\eqref{Eq:model}. Correspondingly
a quantum anomalous Hall conductance appears. To have a paramagnetic
Weyl semimetal, we have to introduce a time reversal counterpart
for the two Weyl nodes in Eq.~(\ref{Eq:model}). A straightforward
extension is as follows
\begin{equation}
H=A(k_{x}\alpha_{x}+k_{y}\alpha_{y})+M(k_{c}^{2}-k^{2})\beta,
\end{equation}
where the Dirac matrices are $\alpha_{x}=\sigma_{x}\otimes\sigma_{x}$,
$\alpha_{y}=\sigma_{x}\otimes\sigma_{y}$, $\beta=\sigma_{0}\otimes\sigma_{z}$.
It contains four Weyl nodes, which are doubly degenerate. The surface
electrons around the $\hat{\mathbf{z}}$ direction consist of two branches with opposite
spins and opposite effective velocities. It will give rise to the
quantum spin Hall effect, compared to the quantum anomalous Hall effect
in a Weyl semimetal of a single pair of nodes. The dispersions of two branches of the 0th bands are $E_{\pm}(k_{z})=\pm(\omega/2-Mk_{c}^{2}+Mk_{z}^{2})$.
In a weak magnetic field, the magnetoconductivity is also linear in
magnetic field, as that for Weyl semimetal in Eq.~(\ref{Eq:model}).
However, for a strong field, the strong field dependent Fermi energy
will give a different field dependence of conductivity as the two 0th
bands shift away from each other as increasing field. Finally it
is worthy pointing out that this kind of semimetals is different from
a simple Dirac semimetal described by a pair of degenerate Dirac cones,
$\pm v\vec{k}\cdot\vec{\sigma}$ (e.g., at the phase transition point
between a 3D topological insulator and trivial insulator), in which
there does not exist the surface states as the two nodes are not separated
in momentum space and the model is topologically marginal. Also, the Zeeman effect may also contribute to a linear contribution to the magnetoconductivity \cite{Burkov15jpcm}. But its sign depends on the g-factor of sample and usually its magnetoconductivity is negative, opposite to that in this work.

\section{Conclusions and discussion \label{Sec:conclusion}}
The key conclusion of the present work
is the positive and linear magnetoconductivity of Weyl semimetals
near the Weyl nodes. The Fermi energy is assumed to cross only the 0th Landau bands of the semimetal at low temperatures ($E_F\ll \omega$ and $k_B T\ll \omega$). The magnetoconductivity
depends on the types of scattering potentials, in which the potential
range is a characteristic parameter. Our conclusion is different from the theoretical
predictions of the positive $B^{2}$ magnetoconductivity in several previous
works \cite{Son13prb,Burkov14prl-chiral}, because where higher Fermi energies ($E_F \gg k_B T$) and $\omega \tau_{k_F}^\text{0,tr}\ll \hbar$ were assumed, therefore there were many more Landau bands on the Fermi surface. Recently, a positive longitudinal magnetoconductivity has been observed in several different candidates of topological semimetal \cite{Li16np,ZhangCL16nc,HuangXC15prx,Xiong15sci,YangXJ15arXiv,ZhangC15arXiv,LiCZ15nc,LiH16nc},
and was claimed to be related to the chiral anomaly. We believe that the linear term is one of the indispensable ingredients in the formula of the magnetoconductivity in the intrinsic Weyl semimetal phase. When more Landau levels come into play as the Fermi energy is shifted from near the Weyl nodes, the magnetoconductivity is expected to deviate from linear.

The result of the magnetoconductivity is based on the Born approximation. When the magnetic length becomes much shorter than the range of the disorder potential, electrons may be scattered by the same impurity for multiple times. The Born approximation contains the correlation of two scattering events by the same impurity \cite{Mahan1990}. In this situation, the validity of the Born approximation was questioned in two dimensions \cite{Raikh93prb,Murzin00pu}.
In three dimensions, it is still unclear whether the correlation of two scattering events in the Born approximation is the building block for the multiple scattering under extremely strong magnetic fields \cite{Song15prb,Pesin15prb}. So far, theories in three dimensions employ the Born approximation, e.g., the quantum linear magnetoresistance \cite{Abrikosov98prb}. The treatment beyond the Born approximation will be a challenging topic for three-dimensional systems under extremely strong magnetic fields.

\section{Acknowledgment}
This work was supported by the Research Grants Council, University
Grants Committee, Hong Kong under Grant No. 17303714 and the Natural
Science Foundation of China under Grant No. 11574127, and in part by the National Science Foundation under Grant No. NSF PHY11-25915.

\appendix
%
%
%
%

\section{Density of states\label{App:DOS}}

\subsection{In the absence of magnetic field}
In the absence of magnetic field, the DOS can be calculated as $N(E)=\sum_{\mathbf{k},i=\pm} \delta({E _i(\mathbf{k})-E})/V$. After the straightforward calculations, the results of DOS are summarized as followings.

If $A^2\ge 2M^2k_c^2$, the results of DOS are given by
\begin{eqnarray}
N(E) &=& \dfrac{E }{4\pi^2AM} \begin{cases}
 \mathcal{D}_1(E), & |E|\le Mk_c^2\\
 \mathcal{D}_2(E), & |E|>Mk_c^2
\end{cases}
\end{eqnarray}
where
\begin{eqnarray}
\mathcal{D}_1(E)= \ln \dfrac{2A\sqrt{M(Mk_c^2+E)} + A^2+2 ME}{2A\sqrt{M(Mk_c^2-E)} + A^2-2 ME},\\
\mathcal{D}_2(E)= \ln \dfrac{2A\sqrt{M(Mk_c^2+E)} + A^2+2 ME}{\sqrt{4 M^2 E^2+A^4-4A^2M^2k_c^2}}.
\end{eqnarray}

If $A^2<2M^2k_c^2$, the DOS is instead given by
\begin{eqnarray}
N(E) = & \dfrac{E }{4\pi^2AM}\begin{cases}
 \mathcal{D}_1(E), & |E|\le A^2/2M\\
\mathcal{D}_1(E)+\mathcal{D}_3(E), & {A^2}/{2M}<|E|\le Mk_c^2 \\
\mathcal{D}_2(E)+\mathcal{D}_4(E), & |E|>Mk_c^2
\end{cases}
\end{eqnarray}
where
\begin{eqnarray}
\mathcal{D}_3(E) =
& \ln \dfrac{A\sqrt{4M^2k_c^2-2{A^2}}+ \sqrt{4M^2E^2 -A^4} }{\sqrt{4A^2M^2k_c^2-A^4-4M^2E^2}},\\
\mathcal{D}_4(E)
= & \ln \dfrac{{A}\sqrt{4M^2k_c^2-2{A^2}}+{\sqrt{4M^2E^2-A^4}}}{\sqrt{{4{M^2}E^2 + A^4 -4A^2M^2k_c^2}}} .
\end{eqnarray}

For a small energy $E\ll Mk_c^2, A^2/2M$ (in both the cases $A^2>2M^2k_c^2$ and $A^2<2M^2k_c^2$), the DOS can be simplified as
\begin{eqnarray}
  N(E) = \dfrac{E^2}{2\pi^2A^2Mk_c}.
\end{eqnarray}

\subsection{In the presence of magnetic field}
In the presence of magnetic field, the DOS can be calculated as $N(E)=(1/2\pi \ell_B^2 L_z)\sum_{\nu,k_z,s=\pm} \delta({E^{\nu s}_{k_z}-E})$.
Summing over $k_z$ and $s=\pm$, $N(E)$ can be written as
\begin{eqnarray}
N({E})
= & \dfrac{|E-\omega/2|}{4\pi^2 \ell_B^2 \sqrt{M}} \Bigg[ \sum\limits_{\nu=0}^{N_+} \dfrac{1}{\mathcal{P}_\nu(E)} \sqrt{\dfrac{1}{Mk_c^2 - \nu\omega + \mathcal{P}_\nu(E)}}\notag \\
& ~~~~~~~~~~~~~~ + \sum\limits_{\nu=0}^{N_-} \dfrac{1}{\mathcal{P}_\nu(E)} \sqrt{\dfrac{1}{Mk_c^2 - \nu\omega - \mathcal{P}_\nu(E)}} \Bigg],
\end{eqnarray}
where $\mathcal{P}_\nu(E)=  \sqrt{(E -{\omega}/{2})^2 - \nu \eta^2}$ and $N_\pm$ are the largest integers for which $Mk_c^2 - \nu \omega \pm \mathcal{P}_\nu(E)$ are positive, respectively.

In the quantum limit, only the $\nu=0$ term is retained and $N(E)$ reduces to
\begin{eqnarray}
N(E) = & \dfrac{1}{4\pi^2 \ell_B^2} \dfrac{1}{\sqrt{M}\sqrt{E+Mk_c^2-\omega/2}},
\end{eqnarray}
which is approximately proportional to the magnetic field.

\section{Lifetimes and transport times}

\subsection{Transport time $\tau^{0,\text{tr}}_{k_F}$}\label{sec:tau-0tr}
The transport time $\tau^{0,\text{tr}}_{k_F}$ at the Fermi surface is calculated as
\begin{eqnarray}
\frac{\hbar}{\tau^{\text{tr}}_{k_F}}
&=& 2\pi \sum_{k_x',k_z'}
\langle |U^{0,0}_{k_x,k_F;k_x',k_z'}|^2 \rangle \delta(E_F-E^0_{k_z'}) (1-\frac{v^z_{0,k_z'}}{v_F}).
\end{eqnarray}
Using $\delta(E_F-E^0_{k_z'})=[\delta(k_F-k_z')+\delta(k_F+k_z')]/\hbar v_F$ and substituting the expressions Eq.~\eqref{U-00} for the scattering matrix elements $\langle |U^{0,0}_{k_x,k_z;k_x',k_z'}|^2 \rangle$,
\begin{eqnarray}
\frac{\hbar}{\tau_{k_F}^{\text{tr}}}
&=& \frac{1}{\hbar v_F} \sum_{k_x',k_z'}
 \int  \frac{d^3\mathbf{q}}{L_xL_z }\langle U^2(\mathbf{q})\rangle_{\text{imp}} e^{- q_\perp^2\ell_B^2/2} \delta(q_x+k_x'-k_x) \notag \\
 & & \times\delta(q_z+k_z'-k_z) [\delta(k_z'+k_F)+\delta(k_z'-k_F)](1-\frac{v^z_{0,k_z'}}{v_F}).\label{Eq1}
\end{eqnarray}
where $q_\perp^2=q_x^2+q_y^2$. Changing the summations over $k_x',k_z'$ to integrals, Eq.~\eqref{Eq1} can be simplified as
\begin{eqnarray}
\frac{\hbar}{\tau_{k_F}^{\text{tr}}}
&=& \frac{2}{\hbar v_F}\int \frac{dq_xdq_y}{(2\pi)^2 }   \langle U^2(q_x,q_y,2k_F)\rangle_{\text{imp}} e^{-q_\perp^2\ell_B^2/2}.
\end{eqnarray}
The substitution of the Fourier transform of the potential $U_i(\mathbf{q}) = u_i  e^{-q^2d^2/2}$ gives
\begin{eqnarray}
\frac{\hbar}{\tau_{k_F}^{\text{tr}}}
&=& \frac{2V_{\text{imp}}}{\hbar v_F}\int \frac{dq_xdq_y}{(2\pi)^2 }e^{-q_\perp^2(d^2+\ell_B^2/2)} e^{- 4d^2 k_F^2 }, \label{tau-0}
\end{eqnarray}
where $V_{\text{imp}}\equiv \sum_i u_i^2/V$. Performing the integral in polar coordinates and we finally obtain
\begin{eqnarray}\label{tau-tr-gaussian-twonode-B}
\frac{\hbar}{\tau_{k_F}^{ \text{tr}}}
&=& \frac{V_{\text{imp}}}{\hbar v_F} \frac{ e^{- 4d^2k_F^2}}{2\pi (d^2+\ell_B^2/2)}.
\end{eqnarray}

\subsection{Lifetimes $\tau_{k_F}^{1\pm}$}\label{sec:tau-0}

We assume that the impurity scattering is so weak that when the Fermi energy crosses only the $0$th landau bands, the leading order of the lifetimes $\tau^{1\pm}_{k_F}$  of the states in the $1\pm$ Landau bands can be found as
\begin{eqnarray}
\frac{\hbar}{\tau_{k_F}^{1\pm}} &=& 2\pi \sum_{k_x',k_z'}
\langle |U^{1\pm,0}_{k_x, k_F;k_x',k_z'}|^2 \rangle \delta(E_F-E^0_{k_z'}).
\end{eqnarray}
Substituting Eq.~(\ref{U-1+}) for the scattering matrix elements $\langle |U^{1\pm,0}_{k_x, k_F;k_x',k_z'}|^2 \rangle$ and then using $\delta(E_F-E^0_{k_z'})=[\delta(k_F-k_z')+\delta(k_F+k_z')]/\hbar v_F$, we have
\begin{eqnarray}\label{tau1p0kF-twonode}
\frac{\hbar}{\tau^{1\pm}_{k_F}}
&=&  \frac{\ell_B^2}{4\hbar v_F} (1\mp \cos{\theta^1_{k_F}})\int_{-\infty}^{\infty} \frac{dq_xdq_y}{ (2\pi)^2 }q_\perp^2
e^{-{\ell_B^2}q_\perp^2/{2}}\notag \\
&& ~~~~~ \times \left[\langle U^2(q_x,q_y,0)\rangle_{\text{imp}}+\langle U^2(q_x,q_y,2k_F)\rangle_{\text{imp}} \right].
\end{eqnarray}
Substituting the Fourier transform of the Gaussian potential and then performing the integrals, Eq.~\eqref{tau1p0kF-twonode} become
\begin{eqnarray}
\frac{\hbar}{\tau_{k_F}^{1\pm}}
&=&\frac{V_{\text{imp}}\ell_B^2}{16\pi\hbar v_F} \frac{1 +e^{- d^2  (2k_F)^2}}{(d^2+\ell_B^2/2)^2}(1\mp \cos{\theta^1_{k_F}}).\label{tau1plus}
\end{eqnarray}
In the strong-field limit $\ell_B\rightarrow 0$, $\cos\theta^1_{k_F} \rightarrow -\text{sgn}(M)$. Suppose $M>0$, then in the strong-field limit, $\hbar/\tau_{k_F}^{1-}=0$ while
\begin{eqnarray}\label{tau-1-gaussian-twonode-strongB}
\frac{\hbar}{\tau_{k_F}^{1+}} &=&\frac{V_{\text{imp}}\ell_B^2}{8\pi\hbar v_F} \frac{1 +e^{- d^2 (2k_F)^2}}{(d^2+\ell_B^2/2)^2}.
\end{eqnarray}

\section{Scattering matrix elements \label{matrixelements}}
To evaluate the square of the scattering matrix elements under the average of impurity configurations, we need to calculate $\sum_{i,j}I_{\nu,\mu}(\mathbf{R}_i)I^*_{\nu',\mu'}(\mathbf{R}_j)$. The sum runs over all impurities. The integrals $I_{\nu,\mu}(\mathbf{R}_i)$ is defined as
\begin{eqnarray}\label{I-def}
I_{\nu,\mu}(\mathbf{R}_i)&=&\frac{1}{L_xL_z} \int d\mathbf{r} \varphi^{\nu*}_{k_x}(y) \varphi^{\mu}_{k_x'}(y) U_i(\mathbf{r}-\mathbf{R}_i) e^{i(k_x'-k_x)x+i(k_z'-k_z) z},
\end{eqnarray}
where $U_i(\mathbf{r}-\mathbf{R}_i)$ denotes the scattering potential of a single impurity at the position $\mathbf{R}_i$. The $\hat{\bf{y}}$-direction wavefunction $\varphi_{k_x}^\nu(y)$ is given in Eq.~\eqref{psi-nukxkz}.  Fourier transforming the potential by using
\begin{eqnarray}
U_i(\mathbf{r}-\mathbf{R}_i)&=&
\int \frac{d\mathbf{q}}{(2\pi)^3} U_i(\mathbf{q})
e^{i\mathbf{q}\cdot (\mathbf{r}-\mathbf{R}_i)},
\end{eqnarray}
and then using the formula $\int_{-\infty}^\infty dx e^{ikx}=2\pi\delta(k)$,
$I_{\nu,\mu}$ can be rewritten as
\begin{eqnarray}
I_{\nu,\mu}(\mathbf{R}_i)
&=& \int \frac{d^3\mathbf{q}}{2\pi L_xL_z} e^{-i\mathbf{q}\cdot \mathbf{R}_i}U_i(\mathbf{q})\int dy \varphi^{\nu*}_{k_x}(y)\varphi^{\mu}_{k_x'}(y) \notag \\
&& \times e^{iq_y y}\delta(q_x+k_x'-k_x) \delta(q_z+k_z'-k_z). \label{funI}
\end{eqnarray}
Using Eq.~\eqref{funI}, $\sum_{i,j}I_{\nu,\mu}(\mathbf{R}_i)I^*_{\nu',\mu'}(\mathbf{R}_j)$ can be written as
\begin{eqnarray}
&& \sum_{i,j}I_{\nu,\mu}(\mathbf{R}_i) I^*_{\nu',\mu'}(\mathbf{R}_j)\nonumber\\
&=&\sum_{i,j} \int \frac{d^3\mathbf{q}}{2\pi L_xL_z} e^{-i\mathbf{q}\cdot \mathbf{R}_i}U_i(\mathbf{q})\delta(q_x+k_x'-k_x)
 \delta(q_z+k_z'-k_z)\nonumber\\
  &&\times \int dy \varphi^{\nu*}_{k_x}(y) \varphi^{\mu}_{k_x'}(y)
e^{iq_yy}\nonumber\\
& &\times  \int \frac{d^3\mathbf{q}'}{2\pi L_xL_z}  e^{i\mathbf{q}'\cdot \mathbf{R}_j}U_j(\mathbf{q}')\delta(q'_x+k_x'-k_x)
 \delta(q_z'+k_z'-k_z)\nonumber\\
&& \times
\int dy' \varphi^{\nu'}_{k_x}(y') \varphi^{\mu'*}_{k_x'}(y')
e^{-iq_y' y'}.
\end{eqnarray}

Under the average of impurity configurations, we employ the theory of impurity average \cite{Mahan1990}
\begin{equation}
 \langle \sum_{i,j} e^{i\mathbf{q}' \cdot \mathbf{R}_j-i \mathbf{q} \cdot \mathbf{R}_i} U_i(\mathbf{q}) U_j(\mathbf{q}')\rangle
 \approx \sum_i U_i^2(\mathbf{q}) \delta_{\mathbf{q}-\mathbf{q}'}
 = (2\pi)^3 \langle U^2(\mathbf{q}) \rangle_{\text{imp}} \delta(\mathbf{q}-\mathbf{q}'),
\end{equation}
where $\langle U^2(\mathbf{q}) \rangle_{\text{imp}}\equiv \sum_i U_i^2(\mathbf{q})/{V}$. Then we can integrate over $\mathbf{q}'$ to obtain
\begin{eqnarray}\label{II-3}
\mathcal{I}_{\nu,\mu}^{\nu',\mu'}&\equiv& \langle \sum_{i,j}I_{\nu,\mu}(\mathbf{R}_i)
I^*_{\nu',\mu'}(\mathbf{R}_j)\rangle_{\text{imp}} \notag \\
&=& \int \frac{d^3\mathbf{q}}{2\pi L_xL_z }\langle U^2(\mathbf{q})\rangle_{\text{imp}}\delta(q_x+k_x'-k_x)
 \delta(q_z+k_z'-k_z)  \nonumber\\
&& \times \int dy \varphi^{\nu*}_{k_x}(y)\varphi^{\mu}_{k_x'}(y)
e^{iq_y y}\int dy'\varphi^{\nu'}_{k_x}(y')\varphi^{\mu'*}_{k_x'}(y')
e^{-iq_y y'},
\end{eqnarray}
where we have used
\begin{eqnarray}
\delta^2(q_x+k_x'-k_x) &=&\frac{L_y}{2\pi}\delta(q_x+k_x'-k_x),\\
\delta^2(q_z+k_z'-k_z) &=&\frac{L_z}{2\pi} \delta(q_z+k_z'-k_z).
\end{eqnarray}

\subsection{Scattering matrix element $\langle
|U^{0,0}_{k_x,k_z;k_x',k_z'}|^2\rangle $}
In the Landau gauge $\mathbf{A}=-By\hat{\mathbf{x}}$, the corresponding eigen wavefunctions for $\nu\ge1$ can be written
\begin{eqnarray}\label{LL-nu}
\langle \mathbf{r}|\nu +,k_{x},k_{z}\rangle & = & \begin{bmatrix}
  \cos\frac{\theta_{k_{z}}^{\nu}}{2}~\psi_{\nu-1,k_x k_z}(\mathbf{r})\\
\sin\frac{\theta_{k_{z}}^{\nu}}{2}~\psi_{\nu,k_x k_z}(\mathbf{r})
\end{bmatrix},\nonumber \\
\langle \mathbf{r}|\nu -,k_{x},k_{z}\rangle & = & \left[\begin{array}{cc}
\sin\frac{\theta_{k_{z}}^{\nu}}{2}~\psi_{\nu-1,k_x k_z}(\mathbf{r})\\
-\cos\frac{\theta_{k_{z}}^{\nu}}{2}~\psi_{\nu,k_x k_z}(\mathbf{r})
\end{array}\right],
\end{eqnarray}
while for $\nu=0$ as
\begin{eqnarray}\label{LL-nu0}
\langle \mathbf{r}|0,k_{x},k_{z}\rangle=\left[\begin{array}{cc}
0\\
\psi_{0,k_x k_z}(\mathbf{r})
\end{array}\right],
\end{eqnarray}
where $\cos\theta^\nu_{k_z}=\mathcal{M}_{\nu}/\sqrt{\mathcal{M}_{\nu}^{2}+\nu\eta^{2}}$. The wavefunctions $\psi_{\nu,k_x k_z}(\mathbf{r})$ are given by
\begin{eqnarray}
\psi_{\nu,k_{x} k_{z}}(\mathbf{r}) &=& \frac{e^{ik_{x}x+ik_{z}z}}{\sqrt{L_{x}L_{z}}}\varphi_{\nu,k_{x}}(y), \notag \\
\varphi_{\nu,k_{x}}(y) &=& \frac{e^{-{(y-y_{0})^{2}}/{2\ell_{B}^{2}}}}{\sqrt{\nu!2^{\nu}\sqrt{\pi}\ell_{B}}} \mathcal{H}_{\nu}(\frac{y-y_{0}}{\ell_{B}}),\label{psi-nukxkz}
\end{eqnarray}
where $y_{0}=k_{x}\ell_{B}^{2}$ is the guiding center and $\mathcal{H}_{\nu}$ are the Hermite polynomials.

Using the wavefunctions Eq.~(\ref{LL-nu}), the scattering matrix elements between state $|0,k_x,k_z\rangle$ and state $|0,k_x',k_z'\rangle$ can be written as
\begin{eqnarray}
U^{0,0}_{k_x,k_x;k_x',k_z'} \equiv & \langle 0,k_x,k_z |U(\mathbf{r})|0,k_x',k_z'\rangle
= \sum_{i}  I_{0,0}(\mathbf{R}_i).\label{Unn1-I}
\end{eqnarray}
and thus $\langle|U^{0,0}_{k_x,k_z;k_x',k_z'}|^2\rangle=\mathcal{I}_{0,0}^{0,0}$. With $\nu=0$ and $\mu=0$ in Eq. (\ref{II-3}), we have
\begin{eqnarray}
\mathcal{I}_{0,0}^{0,0}
 &=& \int \frac{d^3\mathbf{q}}{2\pi L_xL_z}\langle U^2(\mathbf{q})\rangle_{\text{imp}}\delta(q_x+k_x'-k_x) \label{II-00}\\
&&\times \delta(q_z+k_z'-k_z)\Big| \int dy \varphi^{0*}_{k_x}(y) \varphi^{0}_{k_x-q_x}(y)
e^{iq_y y} \Big|^2 .\nonumber
\end{eqnarray}
Substituting the explicit form of the wavefunction Eq. (\ref{psi-nukxkz}), we can find
\begin{eqnarray}
&& \int dy \varphi^{0*}_{k_x}(y)\varphi^{0}_{k_x-q_x}(y)
e^{iq_y y}
 = e^{-{\ell_B^2}(q_\perp^2-2 i q_x q_y + i 4 k_x q_y)/{4}},
\end{eqnarray}
and hence
\begin{eqnarray}\label{integral-0000}
\Big| \int dy \varphi^{0*}_{k_x}(y)\varphi^{0}_{k_x-q_x}(y)
e^{iq_y y} \Big|^2
 =e^{-q_\perp^2{\ell_B^2}/2}.
\end{eqnarray}
Therefore Eq.~\eqref{II-00} becomes
\begin{equation}
\langle|U^{0,0}_{k_x,k_z;k_x',k_z'}|^2\rangle
= \int \frac{d^3\mathbf{q}}{ 2\pi L_xL_z }\langle U^2(\mathbf{q})\rangle_{\text{imp}} e^{-q_\perp^2 \ell_B^2/2}
  \delta(q_x+k_x'-k_x) \delta(q_z+k_z'-k_z).\label{U-00}
\end{equation}

\subsection{Scattering matrix element $\langle |U^{1\pm,0}_{k_x,k_z;k_x',k_z'}|^2\rangle $}
Similarly, the scattering matrix elements between state $|1\pm,k_x,k_z\rangle$ and state $|0,k_x',k_z'\rangle$ can be written as
\begin{eqnarray}
U^{1+,0}_{k_x,k_z;k_x',k_z'} &\equiv & \langle 1+,k_x,k_z |U(\mathbf{r}) |0,k_x',k_z'\rangle= \sum_{i} \sin\frac{\theta^1_{k_z}}{2} I_{1, 0}(\mathbf{R}_i), \label{Unn1-I} \\
U^{1-,0}_{k_x,k_z;k_x',k_z'} &\equiv & \langle 1-,k_x,k_z |U(\mathbf{r}) |0,k_x',k_z'\rangle = \sum_{i} \cos\frac{\theta^1_{k_z}}{2} I_{1, 0}(\mathbf{R}_i),
\end{eqnarray}
and thus $\langle|U^{1\pm,0}_{k_x,k_z;k_x',k_z'}|^2\rangle = \mathcal{I}_{1,0}^{1,0} ( 1\mp\cos \theta^1_{k_z})/2$.
With $\nu=1$ and $\mu=0$ in Eq. (\ref{II-3}), we have
\begin{eqnarray}
\mathcal{I}_{1,0}^{1,0}
&=& \int \frac{d^3\mathbf{q}}{2\pi L_xL_z }\langle U^2(\mathbf{q})\rangle_{\text{imp}}\Big| \int dy \varphi^{1*}_{k_x}(y)\varphi^{0}_{k_x'}(y)
e^{iq_y y} \Big|^2 \nonumber\\
&& \times\delta(q_x+k_x'-k_x)
 \delta(q_z+k_z'-k_z) . \label{II-10}
\end{eqnarray}
Substituting the wavefunction Eq. (\ref{psi-nukxkz}), we can find
\begin{eqnarray}
& & \int dy \varphi^{1*}_{k_x}(y)\varphi^{0}_{k_x-q_x}(y)
e^{iq_y y}
= -\frac{\ell_B(q_x+i q_y)}{\sqrt{2}}
  e^{-{\ell_B^2}(q_\perp^2-2 i q_x q_y+i 4 k_x q_y)/4},
\end{eqnarray}
and hence
\begin{eqnarray}\label{integral-no-0q-limit}
 \Big| \int dy \varphi^{1*}_{k_x}(y)\varphi^{0}_{k_x-q_x}(y) e^{iq_y y} \Big|^2
&=& \dfrac{q_\perp^2\ell_B^2}{2}e^{-q_\perp^2\ell_B^2/2}.
\end{eqnarray}
Therefore Eq.~\eqref{II-10} can be rewritten as
\begin{equation}\label{II-4}
\mathcal{I}_{1,0}^{1,0}
= \frac{\ell_B^2}{2} \int \frac{d^3\mathbf{q}}{ 2\pi L_xL_z } \langle U^2(\mathbf{q})\rangle_{\text{imp}} q_\perp^2
e^{-q_\perp^2\ell_B^2/2}
 \delta(q_x+k_x'-k_x) \delta(q_z+k_z'-k_z),
\end{equation}
and
\begin{eqnarray}\label{U-1+}
 \langle|U^{1\pm,0}_{k_x,k_z;k_x',k_z'}|^2\rangle
&=& \frac{\ell_B^2}{4} ( 1\mp\cos \theta^1_{k_z}) \int \frac{d^3\mathbf{q}}{ 2\pi L_xL_z }\langle U^2(\mathbf{q}) \rangle_{\text{imp}}q_\perp^2 e^{-q_\perp^2{\ell_B^2}/2} \notag \\
&& \times \delta(q_x+k_x'-k_x) \delta(q_z+k_z'-k_z).
\end{eqnarray}


%

\end{document}